\documentclass{iopconfser}
\usepackage{graphicx} 

\usepackage{xcolor}
\usepackage[breaklinks=true,colorlinks=true,linkcolor=blue,urlcolor=blue,citecolor=blue]{hyperref}
\usepackage{multirow, booktabs} 
\usepackage{tabularx}
\usepackage{orcidlink}
\usepackage{array}
\newcommand{\PreserveBackslash}[1]{\let\temp=\\#1\let\\=\temp}
\newcolumntype{C}[1]{>{\PreserveBackslash\centering}p{#1}}
\newcolumntype{R}[1]{>{\PreserveBackslash\raggedleft}p{#1}}
\newcolumntype{L}[1]{>{\PreserveBackslash\raggedright}p{#1}}
\usepackage{doi}
\usepackage{lmodern}
\begin{document}
\title{Improved accuracy of the  reference photon calibration standards in the network of gravitational-wave observatories during the O4 observing run}

\author{D.~Bhattacharjee$^{1,2}$\,\orcidlink{0000-0001-6623-9506},  
C.~Grimaud$^3$\,\orcidlink{0000-0001-7736-7730},  
M.~Spidell$^4$\,\orcidlink{0000-0002-5513-8520},
S.~Kück$^5$\,\orcidlink{0000-0003-2188-483X}, 
D.~Chen$^6$\,\orcidlink{0000-0003-1433-0716},
J.~Betzwieser$^7$\,\orcidlink{0000-0003-1533-9229},
S.~Hido$^8$\,\orcidlink{0009-0009-0004-4170}, 
P.~Lagabbe$^9$\,\orcidlink{0009-0004-9454-3125},
H.~Lecher$^5$\orcidlink{0009-0003-0853-0508},
J.~Lehman$^4$\,\orcidlink{0000-0003-4729-7123},
F.~Llamas$^{10}$\,\orcidlink{0009-0008-4146-000X},
M.~López$^5$\orcidlink{0000-0002-0706-1418},
L.~Rolland$^3$\,\orcidlink{0000-0003-0589-9687},
A.~Sanchez$^{11}$\,\orcidlink{0009-0004-7716-1061},  and
R.~L.~Savage$^{11}$\,\orcidlink{0000-0003-3317-1036}
}  
\affil{$^1$Kenyon College, Gambier, USA}
\affil{$^2$Missouri University of Science and Technology, Rolla, USA}
\affil{$^3$Université Savoie Mont Blanc, CNRS, Laboratoire d’Annecy de Physique des Particules
- IN2P3, F-74000 Annecy, France}
\affil{$^4$National Institute of Standards and Technology, Boulder, USA}
\affil{$^5$Physikalisch-Technische Bundesanstalt, Braunschweig, Germany}
\affil{$^6$National Astronomical Observatory of Japan, Hida, Japan}
\affil{$^7$LIGO Livingston Observatory, Livingston, USA}
\affil{$^8$Institute for Cosmic Ray Research, Kashiwa, Japan}
\affil{$^{9}$European Gravitational Observatory, Cascina, Italy}
\affil{$^{10}$University of Texas Rio Grande Valley, Brownsville, USA}
\affil{$^{11}$LIGO Hanford Observatory, Richland, USA}

\email{driptab@caltech.edu\,\footnote{Current affiliation: LIGO Livingston Observatory, Livingston, USA}, rsavage@caltech.edu}

\begin{abstract}
The global network of gravitational wave observatories uses systems known as Photon Calibrators to provide absolute calibration via laser radiation pressure. 
Successive detector improvements are increasing sensitivity, enabling not only increase the detection rate but also detections with high signal to noise ratios.
To optimally extract information from these detected signals, continual effort is directed at pushing the state of the art of the interferometer calibration accuracy.
To improve the accuracy of the locally-maintained laser power standards for photon calibration, a new global calibration scheme was implemented by using transfer standards calibrated at both the National Institute of Standards and Technology (NIST) in the US and the Physikalisch-Technische Bundesanstalt (PTB) in Germany during the recently completed fourth observing run. 
We report the results of a long-duration NIST-PTB bilateral comparison conducted over almost three years and the results of propagation of transfer standard calibrations to the reference power sensors maintained at each observatory.
\end{abstract}

\section{Introduction}
\label{sec:Int}

The two detectors of the Laser Interferometer Gravitational-wave Observatory (LIGO) \cite{detectorpaper, O4LIGO} located in the USA, the detectors of the Virgo Project \cite{AdvVirgoPaper, O4Virgo} in Italy and the KAGRA Project\cite{kagradet} in Japan, constitute the global network of ground-based gravitational wave (GW) detectors.
Due to increasing detector sensitivity and detection rates, eventually calibration of the interferometer
output signals with sub-percent accuracy will be required for optimal extraction of the astrophysical information embedded within the signals. 
This necessitates providing absolute calibration of displacement fiducials at well below the sub-percent level.

Current gravitational-wave interferometers use systems that are referred  to as Photon Calibrators (Pcals) to provide absolute calibration. 
The Pcals produce periodic  fiducial displacements of suspended interferometer mirrors at discrete frequencies across the 20 Hz to 2 kHz detection band via photon radiation pressure~\cite{RSIpaper, virgopcal2, DanPcal}. 
Since these periodic displacements of the mirrors are proportional to the incident Pcal laser power, accurate and precise calibration of the displacement fiducials depends on accurate estimation of the power impinging on the mirrors. 

To date, LIGO Pcal power sensor calibrations have been based on transfer standards (TS) calibrated by the US national metrology institute (NMI), the National Institute of Standards and Technology (NIST) \,\cite{RSIpaper, O3PcalPaper}.
Virgo and KAGRA Pcal power standards were also calibrated by LIGO using a NIST-calibrated TS.
This enabled identification and elimination of $\sim$ 4\,\% relative error between the LIGO and Virgo detectors at the beginning of a previous observing run~\cite{virgopcal2}.
In order to further improve confidence in calibration accuracy and to reduce relative calibration errors between observatories, a new {\em global} calibration scheme \cite{NEWRAD}, involving both NIST and the Physikalisch-Technische Bundesanstalt (PTB) in Germany, was implemented during the two and a half years of the O4 observing run, from May, 2023 to November, 2025.

Calibrating TSs at NIST and PTB and propagating these NMI calibrations to the local Pcal reference sensors maintained at each observatory are the first two steps in the global calibration scheme and the main focus of  this paper.
A detailed description of the subsequent steps for calibrating Pcal-induced displacement fiducials and their uncertainty estimates is provided in~\cite{DanPcal, NEWRAD, Cervane_PhdReport}.
In this paper, the results of the NIST and PTB calibration of the TSs are given in Section \ref{sec:NIST_PTB_study}. 
The results of propagating these NMI calibrations to the local Pcal reference sensors are reported in Section \ref{sec:GScalib}. 
Conclusions are presented in Section \ref{sec:conclusion}.

\section{Long-duration NIST-PTB bilateral comparison }
\label{sec:NIST_PTB_study}

More than 200 confirmed GW signals were detected during the O4 run, all calibrated using NIST and PTB calibrations of the LIGO Pcal power standards. 
A typical LIGO Pcal power sensor, as well the TSs TSA and TSB, comprise an integrating sphere with a Spectralon\textsuperscript{\textregistered} interior shell and custom-built, unbiased InGaAs photodetector.

\begin{figure}[ht]%
\begin{center}
\includegraphics[trim= 0cm 3cm 0cm 3cm, clip=true, width = 0.8\textwidth]{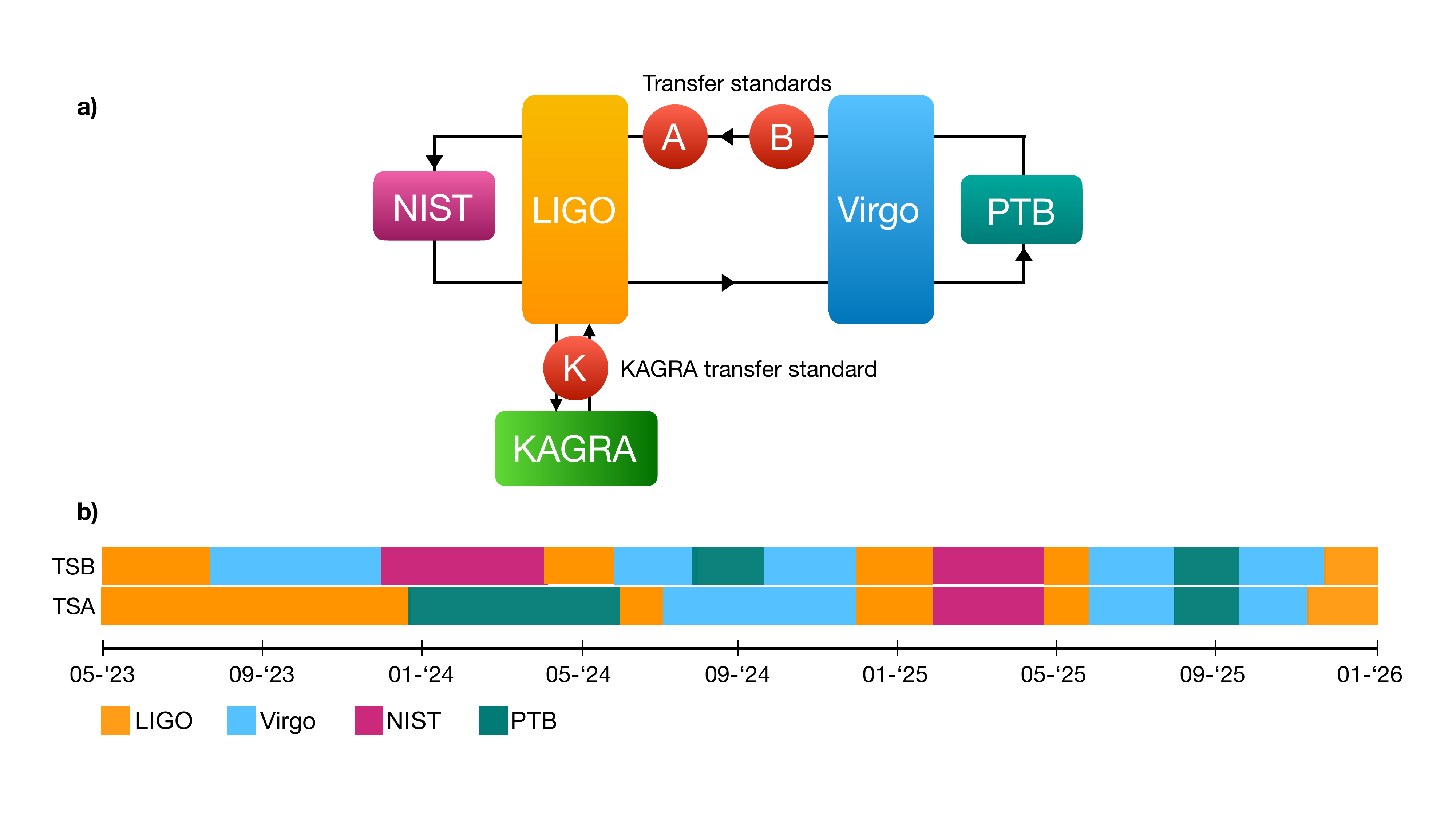}
\caption{{\bf a.} 
Schematic diagram of the global Pcal calibration scheme. Transfer standards TSA and TSB travel around the loop such that each observatory received a NMI-calibrated transfer standard periodically throughout O4. 
{\bf b.} 
A timeline of the locations of TSA and TSB. }
\label{fig:Pcalglobal}
\end{center}
\end{figure}

TSA and TSB, were calibrated at NIST in October 2022, then at PTB, and finally again at NIST, to initiate the implementation of the global calibration scheme shown schematically in Fig.\,\ref{fig:Pcalglobal}. 
The bilateral degree of equivalence (DoE) was \ - 0.20\,\% with expanded relative uncertainty of 0.32\,\%~(k = 2)\cite{NEWRAD}.
This indicated that the scale representation realized by both NMIs were consistent.
The NIST-PTB consensus responsivities of TSA and TSB were used to calibrate the Pcal power sensors at each observatory at the start of the O4 run.
The results of this initial NIST-PTB bilateral study and propagating the NMI calibrations to the LIGO and Virgo Pcal power sensors were published in\,\cite{NEWRAD}. 

\begin{figure}[ht]%
\begin{center}
\includegraphics[trim= 0cm 0cm 0cm 0cm, clip=true, width = 0.75\textwidth]{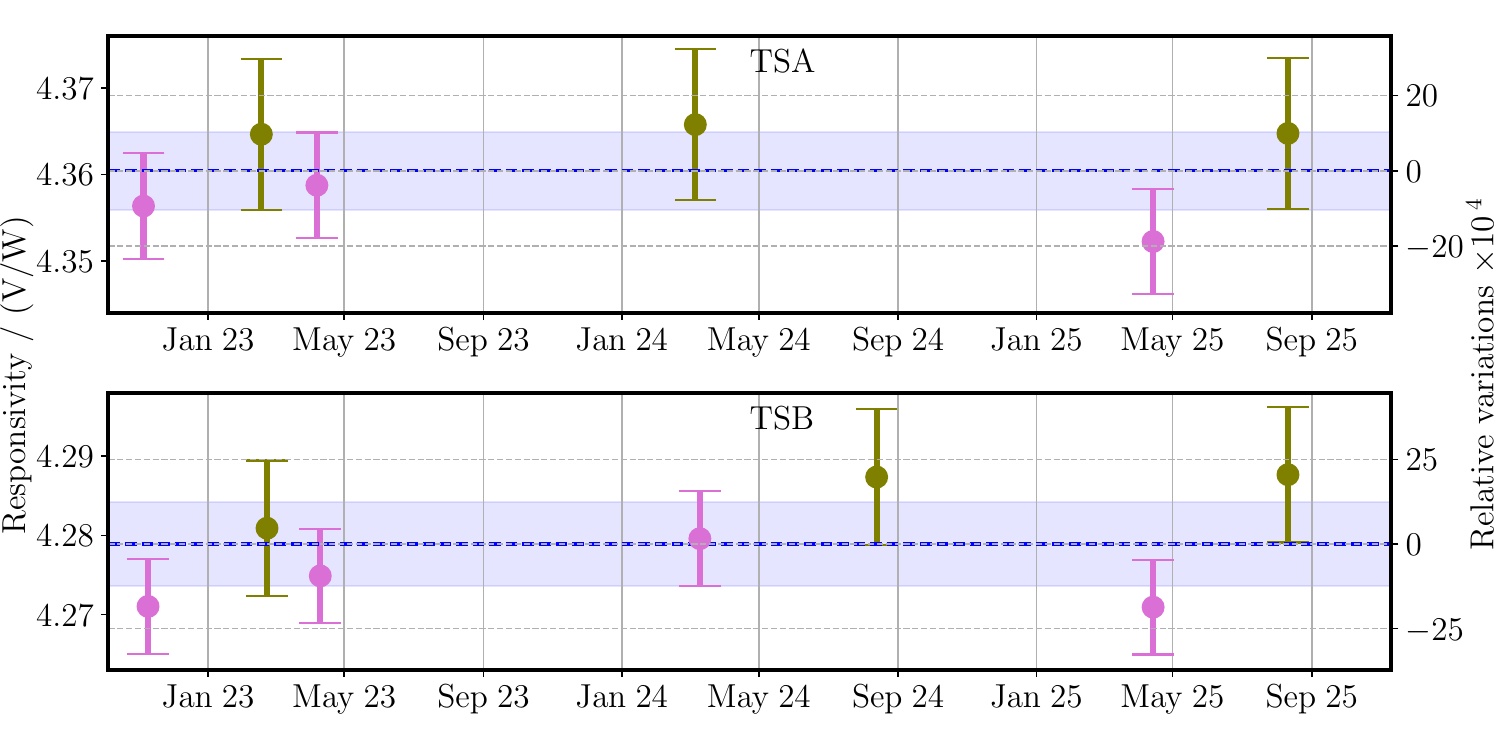}
\caption{ NIST (pink) and PTB (green) measurements of the TSA (top) and TSB (bottom) responsivity over a time span of 3 years.
The error bars are the expanded uncertainty (k=2) of 0.14\,\% and 0.20\,\% for NIST and PTB respectively. The dashed blue line represents the mean of all NIST and PTB measurements. The colored band in each panel represents $\pm$ twice the  relative standard error on the mean. }
\label{fig:meanTScalibrations}
\end{center}
\end{figure}

During the O4 run, TSA and TSB were periodically calibrated at NIST and PTB, resulting in a long-duration NIST-PTB bilateral study. 
The calibration results are shown in Fig.\,\ref{fig:meanTScalibrations}. 
All measurements were made with 300\,mW of 1047 nm laser radiation. 
Since the NMI laboratories are maintained at different temperatures, the measurements are corrected to a reference temperature of 300.15\,K using measured temperature coefficients for each sensor. 
The TSA and TSB responsivities and their uncertainties are tabulated in Table \ref{tab:KeyResults}. 
Long-duration NIST-PTB measurements show that the power standards are stable and reproducible within the level of the peak-to-peak variations of $\pm$ 0.20\,\% for TSA and $\pm$ 0.25\,\% for TSB.

LIGO and Virgo received an NMI-calibrated TS every few months.
The observatories performed responsivity ratio measurements in dedicated optics laboratories to transfer the NMI-calibration to a reference Pcal power standard, a {\em Gold Standard} (GS), maintained at each observatory, before shipping the TS to its next destination.
The timeline of the locations of TSA and TSB during the O4 run is shown in Fig.\,\ref{fig:Pcalglobal}.
Since KAGRA joined the calibration scheme later, a separate TS for KAGRA (TSK), was calibrated using the LIGO GS (GSL) as a reference and shipped to the University of Toyama.
TSK was used to calibrate the KAGRA GS (GSK) before they joined LIGO and Virgo for the O4 run.  

\begin{table}[ht]
\caption{Key results -  TSA and TSB responsivities, and LIGO, Virgo and KAGRA GS responsivities along with their respective statistical uncertainty (rel. unc.)  - for the O4 global calibration scheme. The number of measurements, n, used to estimate the relative standard error for each power standard is also indicated.}
\vspace{0.1in}
\begin{tabular}{ C{3cm} C{3cm} C{2.5cm} C{3.5cm}}
    \toprule[1.0pt]
    \bf{Sensor} & \bf Responsivity /(V/W) &  \bf rel. unc./\%  & Type \\
    \midrule[0.2pt]
     TSA &  $4.3592$ & 0.05 & $u_{rel, \, \rm{Type\,A}} (n=6)$\\ 
     TSB & $4.2789$ & 0.06 &
     $u_{rel, \, \rm{Type\,A}} (n=7)$\\ 
     \midrule[0.2pt]
     GSL & $5.1349$ & 0.01 & $u_{rel, \, \rm{Type\,A}} (n=23)$ \\
     GSV & $2.6012$ & 0.04 & $u_{rel, \, \rm{Type\,A}} (n=9)$ \\
     GSK & 4.7010 & 0.01 & $u_{rel, \, \rm{Type\,A}} (n=5)$  \\
    \bottomrule[1.0pt] 
\\
\end{tabular}
\label{tab:KeyResults}
\end{table}

\section{Calibrating Gold standards at each observatory}
\label{sec:GScalib}

\begin{figure}[ht]%
\begin{center}
\includegraphics[trim= 0cm 0cm 0cm 0cm, clip=true, width = 0.75\textwidth]{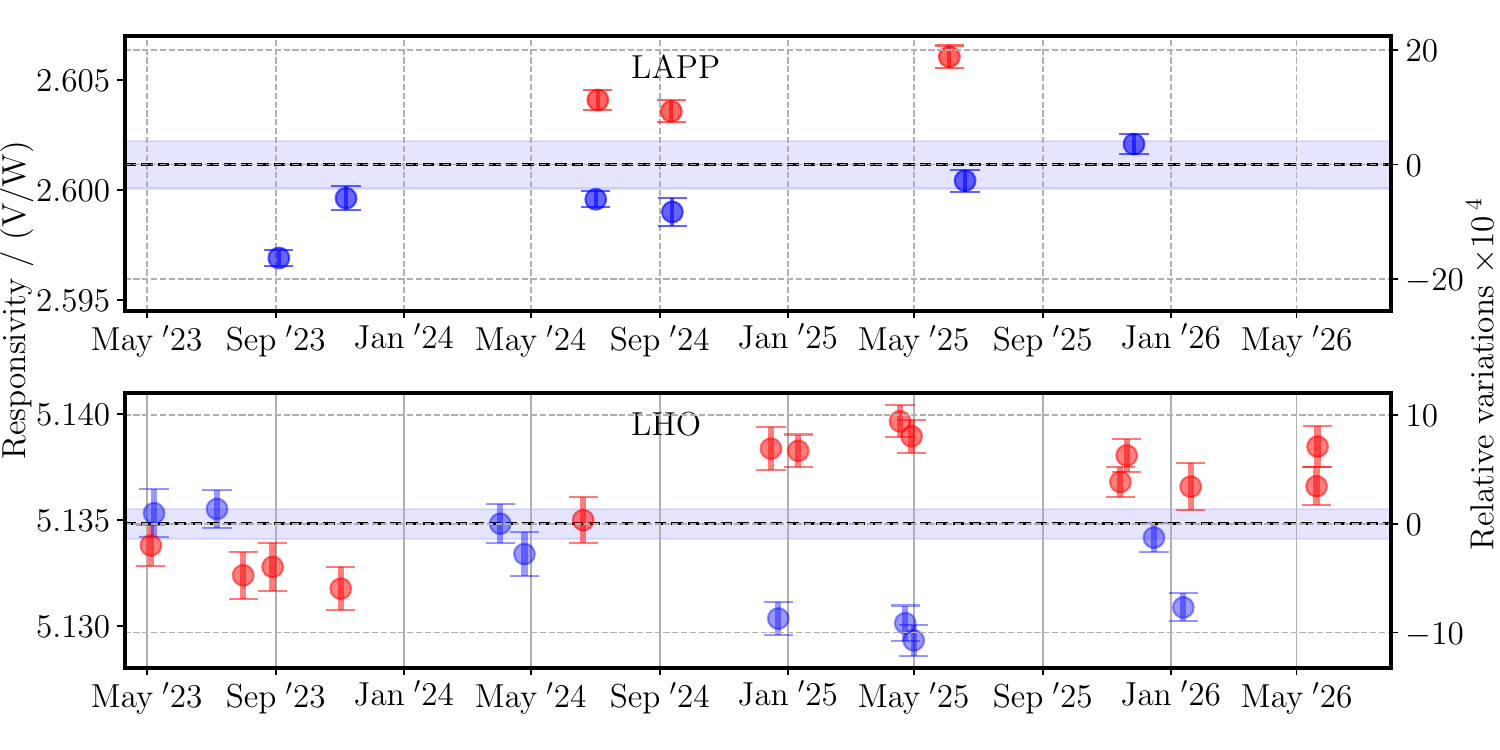}
\caption{Responsivity of the Virgo GS (top panel) and LIGO GS (bottom panel) in (V/W) using  NMI-calibrated TSA (red) and TSB (blue), computed from responsivity ratio measurements performed at LAPP and LHO respectively. 
The error bars are the $\pm 1-\sigma$ statistical measurement uncertainties, magnified by a factor of 5 and 50 for LHO and LAPP respectively for better visibility.
The colored band represents the relative standard error of all the measurements performed over a time span of three years to propagate the NMI calibration to the local GS at the observatory. }
\label{fig:rhoPrimeGSmean}
\end{center}
\end{figure}

After receiving a recently calibrated TSA or TSB, resposnsivity ratio measurements were performed in dedicated optics laboratories at each observatory. 
Figure \ref{fig:rhoPrimeGSmean} shows the responsivity of the GS computed from the GS/TS responsivity ratio measurements performed at the Laboratoire d’Annecy de Physique des Particules (LAPP) and the LIGO Hanford Observatory (LHO) to calibrate the GSs. 
Each data point is the mean of repeated responsivity ratio measurement suites, multiplied by the mean responsivity of TSA or TSB. 
The error bars, magnified by a factor of 5 for LHO and 50 for LAPP for better visibility, represent the $\pm 1-\sigma$ statistical variations of each set of measurement suites.
The measurements results over the last 3 years or so have shown that the power standard responsivities are stationary within $\pm$ 0.10\,\% for LIGO and $ \pm$ 0.19\,\% for Virgo.
The shaded region in Fig. \ref{fig:rhoPrimeGSmean} represents $\pm 1$ relative standard error for all of the measurements. The variations in the data are much larger than what would be expected based on measurement uncertainty, indicating that there are unknown systematic uncertainties in the measurement processes at both LHO and LAPP.
KAGRA calibrated their local GS in a laboratory at the University of Toyama using TSK and the results are reported in \cite{DanPcal}. 

The GS responsivities and the estimated uncertainties for each observatory are tabulated in Table~\,\ref{tab:KeyResults}.   
The GS uncertainties during O4 are almost a factor of 3 smaller than the 0.32\,\% from the previous observing run\,\cite{O3PcalPaper}.
This is mostly due to the new PARRoT detector being used as the primary standard at NIST \cite{SPIE}.

\section{Conclusions}
\label{sec:conclusion}

The global calibration scheme that was implemented during the O4 observing run increased confidence in the calibration of transfer standards by NIST and PTB and improved the accuracy of the calibrations of the GSs that are maintained at each GW observatory.  
The long-duration NIST-PTB bilateral comparison that started before the beginning of the O4 and continued until after its conclusion, demonstrated agreement between the NMI calibrations within the stated uncertainties of approximately 0.2\,\%.  
Repeated transfer of the TS calibrations to GSs at the observatories were stationary at the $\pm$ 0.2\,\% peak-to-peak  at LAPP and $\pm$ 0.1\,\% peak-to-peak at LIGO over the three year span.

\begin{figure}[ht]%
\begin{center}
\includegraphics[trim= 2cm 10cm 2cm 9cm, clip=true, width = 0.7\textwidth]{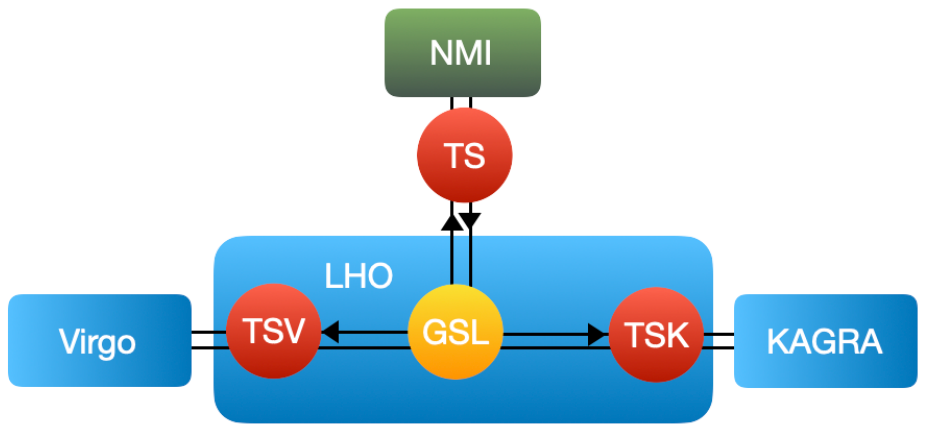}
\caption{Revised global Pcal calibration scheme for future observing runs. A TS travels between NIST and the LHO where the NMI calibration is propagated to GSL. Dedicated TSs for Virgo and KAGRA, TSV and TSK, are calibrated using the GSL as a reference standard at LHO.   }
\label{fig:O5Pcalscheme}
\end{center}
\end{figure}

With the increased confidence in the accuracy and stability of the power sensor calibrations, we plan to calibrate TSs at only one NMI during the next observing run to simplify the global scheme and conserve resources.  This plan is shown schematically in Fig. \ref{fig:O5Pcalscheme}.  An NMI-calibrated TS will be used to calibrate a reference GS maintained at LHO, as well as TSV and TSK that will circulate periodically to the Virgo and KAGRA observatories.

Currently, Pcal and overall calibration uncertainties do not limit accurate parameter estimation from GW signals \cite{VitaleCal, O4uncert}.  However, planning for the next generation of GW observatories such as Einstein Telescope and Cosmic Explorer calls for a ten-fold increase in sensitivity, yielding SNRs as high as 1000 for detected events.  Optimizing extraction of information from these signals will require significant improvements in calibration accuracy.   

The {\em Newtonian Calibrators} (Ncals) that have been developed and implemented by the Virgo project have achieved accuracies similar to the Pcal systems and provide continuous displacement fiducials \cite{NcalO4}.
There is an Ncal prototype already installed at LHO, although it was not used to provide calibrated displacement fiducials during O4.
The KAGRA project is currently installing Ncal systems.
Comparing the fiducial displacements of the suspended mirrors induced by Pcal and Ncal systems should help identify and correct for systematic errors in these systems and thus reduce overall displacement calibration uncertainty.
Ultimately, {\em astrophysical calibration}, using general relativity and the signals themselves, perhaps in conjunction with other messengers \cite{astrocal, GW170817} may play the key role in achieving the level of calibration accuracy required.

\section*{Acknowledgements}

This material is based upon work supported by NSF’s LIGO Laboratory, which is a major facility fully funded by the National Science Foundation (NSF).
The authors are grateful for computational resources provided by the LIGO Laboratory and supported by NSF grants PHY-0757058 and PHY-0823459.  
CG acknowledges support from the French Agence Nationale de la Recherche for the project ACALCO (ANR-21-CE31-0024).
DB and FL acknowledge the LSC Fellows program for supporting their research at LHO under NSF award PHY-1912598. DB acknowledges support from NSF awards OAC-2103662, PHY-2011334, and PHY-2308693. 

This paper carries LIGO document number LIGO-P2600578.

\bibliographystyle{iopart-num-long.bst}
\bibliography{globalcalib}
\end{document}